\documentstyle[12pt,epsf]{article}

\newcommand{\lapprox}{\raisebox{-.2ex}{$\stackrel{\textstyle<}
{\raisebox{-.6ex}[0ex][0ex]{$\sim$}}$}}

\begin{document}

\begin{flushright}
CEBAF-TH-96-06 \\
Corrected
\end{flushright}

\begin{center}
{\Large \bf  Asymmetric Gluon Distributions
and Hard Diffractive Electroproduction}
\end{center}
\begin{center}
{A.V. RADYUSHKIN\footnotemark }  \\
{\em Physics Department, Old Dominion University,}
\\{\em Norfolk, VA 23529, USA}
 \\ {\em and} \\
{\em Thomas Jefferson National  Accelerator Facility,} \\
 {\em Newport News,VA 23606, USA}
\end{center}
\vspace{2cm}

\footnotetext{Also Laboratory of Theoretical Physics, 
JINR, Dubna, Russian Federation}

\begin{abstract}

Due to  the momentum transfer $r \equiv p - p'$ 
from the initial proton to the final,
the ``asymmetric'' matrix element 
$\langle p' | G \ldots G |p \rangle$
that appears in the pQCD description of
hard  diffractive electroproduction
does not coincide with  that defining 
the gluon distribution function $f_g(x)$.
I   outline a lowest-twist pQCD   formalism 
based  on the  concept of   double distribution
$F_g(x,y)$, which specifies  the fractions 
$xp$, $yr$, $(1-y)r$ of the (lightlike) 
initial  momentum $p$ 
and the momentum transfer $r$, $resp.,$ 
carried by the gluons. 
I  discuss    one-loop evolution
equation for the double  distribution
$F_g(x,y;\mu)$ and  obtain  the solution of this  
equation in a simplified situation when the 
quark-gluon mixing effects are ignored. 
For $r^2=0$,  the  momentum transfer  
$r$ is proportional to $p$: $r = \zeta p$, and 
it is  convenient to parameterize
the   matrix element 
$\langle p-r | G \ldots G |p \rangle$
by an asymmetric distribution function
${\cal F}_{\zeta}^g (X)$  depending 
on  the total fractions
$X \equiv x+y \zeta$ and $X-\zeta = x- (1-y) \zeta$   
of  the initial proton momentum $p$
carried by the gluons.   
I formulate evolution 
equations for  ${\cal F}_{\zeta}^g (X)$, study some 
of their general properties and discuss the relationship 
between ${\cal F}_{\zeta}^g (X)$,  $F_g(x,y)$ and $f_g(x)$.

\end{abstract}

\newpage

 {\it 1. Introduction.}  
As shown in ref.\cite{bfgms}, at high  virtualities
of the virtual  photon $\gamma^*$,
one can apply  pQCD factorization
to study 
the process of hard diffractive
exclusive electroproduction of vector mesons 
$\gamma^* + p \to V + p'$.  
In the approach of ref.\cite{bfgms},  the non-perturbative information 
related to the proton is described by the matrix
element of a two-gluon operator
approximated by the gluon distribution function
$f_{g}(x)$. 
However,  as noted in ref.\cite{afs},
due to  the momentum transfer $r \equiv p - p'$ 
from the initial proton to the final,
the two gluons carry, in fact,  different fractions of the
original proton momentum, $i.e.,$
the matrix element of the gluonic operator in this case
does not  coincide with 
that defining  the gluon distribution 
function. As emphasized by X.Ji \cite{ji},
one should deal in this case  with a new type of functions
(he calls them ``off-forward parton distributions'')
which differ from $f_{g}(x)$, even if   $t \equiv r^2$  vanishes. 
My goal in this letter is to develop
a modified   pQCD   approach 
 for the  hard  diffractive
 electroproduction, which takes into account 
the effects related to the momentum transfer.   
In  this formalism,   the basic function 
describing the gluon content of the  ``asymmetric''
matrix element  $\langle p-r | \ldots |p \rangle$
is  the  {\it double
 distribution} 
$F_g(x,y)$, which specifies  the 
fractions $xp$, $yr$, $(1-y)r$ of the 
initial proton momentum $p$ 
and the momentum transfer $r$, $resp.,$ 
carried by the gluons\footnote{Originally, 
the double distributions for
the asymmetric matrix elements
of quark operators were introduced 
in ref.\cite{compton} in application to virtual Compton scattering.}. 
With respect to $x$, the function  $F_g(x,y)$
looks  like a distribution function while 
with respect to $y$  it behaves like a distribution amplitude.
Since the logarithmic scaling violation  is an important 
feature of the gluon distribution function in the low-$x$ region,
I  discuss   the evolution
equation for the double  distribution
$F_g(x,y;\mu)$. The relevant evolution
kernel $R_{gg}(x,y;\xi,\eta)$ 
 produces  the GLAPD evolution kernel $P_{gg}(x/\xi)$ \cite{gl,ap,d}
when  integrated  over $y$, while 
 integrating  $R_{gg}(x,y;\xi,\eta)$ over $x$
gives the expression coinciding with the 
 evolution kernel $V_{gg}(y,\eta)$ 
for the gluon distribution 
amplitude.  
I construct the solution of the 
one-loop evolution equation for the 
double   gluon distribution in a simplified
situation when quark-gluon mixing effects are neglected. 
For  $t=0$ and vanishing hadron masses, the  momentum transfer  
$r$ is proportional to $p$: $r = \zeta p$
and, for this reason,  it is  convenient to parameterize
the  matrix element $\langle p-r | G \ldots G |p \rangle$
by the {\it asymmetric distribution function}
${\cal F}_{\zeta}^g (X)$  specifying the total fractions
$Xp$, $(X-\zeta)p$   of  the initial hadron momentum $p$
carried by the gluons\footnote{The
 asymmetric  distribution functions 
are  similar  to, but not  identical with
the $t \to 0$ limit of the
off-forward parton distributions 
introduced recently by X.Ji \cite{ji}.}.  I formulate 
equations governing the evolution
of the asymmetric distribution function 
${\cal F}_{\zeta}^g (X)$
and discuss the relationship between this function,  
 the double gluon distribution $F_g(x,y)$ 
and the usual gluon distribution function $f_g(x)$.

{\it 2. Double distributions.} 
The  amplitude for the 
elastic  electroproduction process 
$\gamma^* p \to p' V$ 
depends on   the momentum $p$ of the initial proton,
the   momentum  transfer $r=p-p'$ and  the 
momentum $q$  of the produced vector meson.
We will consider the  limit in which one  can neglect 
 the squares of the meson  $q^2 \equiv m_V^2$
and proton $p^2 \equiv m_p^2$ masses compared to
the  virtuality $-Q^2 \equiv (q-r)^2$ of the initial 
photon and the energy invariant $p \cdot q$. 
Thus,   we set  $p^2=0$ and $q^2 = 0$,
and use $q$ and $p$ as  the basic Sudakov light-cone  4-vectors.
In the diffractive region, the invariant momentum  
transfer $r^2 \equiv t$ is  small,  
and we  actually will analyze   the limit  $t=0$. 
Note that  in this case  the  on-shell condition
$p'^2 \equiv (p-r)^2=p^2$  results in the requirement
$(p\cdot r )= 0$  which can be satisfied only
if the  two lightlike momenta
$p$ and $r$ are proportional to each other: $r= \zeta  p$,
where $\zeta \equiv Q^2 / 2 (p \cdot q)$
is the Bjorken variable which obeys $0 \leq \zeta  \leq 1$.

\begin{figure}[ht]
\mbox{
   \epsfxsize=15cm
 \epsfysize=5cm
 \hspace{2cm}  
 \epsffile{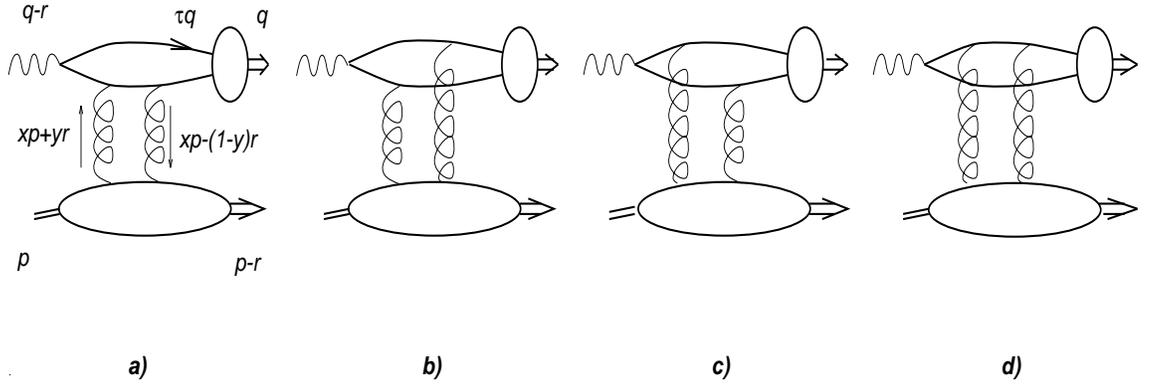}  }
{\caption{\label{fig:1}
 Diagrams contributing 
 to hard diffractive electroproduction
of vector mesons.
   }
}
\end{figure}

The leading contribution in the  large-$Q^2$, fixed-$\zeta $ limit
is given by the 
diagrams shown in  Fig.\ref{fig:1}.
The long-distance 
dynamics is described  there by the  vector meson
distribution amplitude $\varphi_V(\tau)$
and the asymmetric    matrix element 
of the light-cone gluonic operator  
\begin{equation} 
 \langle p -r  \,  | \,   
 G_{\mu \alpha}^a (z_1) E_{ab}(z_1,z_2;A) 
G_{\nu \alpha}^b (z_2)\, | \,p  \rangle  
\label{eq:my}
\end{equation}
in which the 4-vectors $z_1,z_2$ specifying   the
location of  the two
gluon vertices are  separated by 
lightlike intervals $z_1^2=0$, $z_2^2=0$ from the
virtual photon vertex.
The factor $E_{ab}(z_1,z_2;A)$ is the usual
$P$-exponential of the gluonic $A$-field  
along the straight line  connecting $z_1$ and $z_2$,
and the indices $a,b = 1, \ldots 8$ 
denote the gluon color.

The gluon momentum in this matrix element 
originates both from the initial hadron momentum $p$ 
and from the momentum transfer $r$, 
so I write it as $xp+yr$
parameterizing  the asymmetric matrix element of the 
light-cone gluon operator by the double distribution 
$F_g(x,y)$
\begin{eqnarray} 
\hspace{-1cm} &&\lefteqn{\langle p -r  \,  | \,   
z_{\mu}  z_{\nu} G_{\mu \alpha}^a (0) E_{ab}(0,z;A) 
G_{ \alpha \nu}^b (z)\, | \,p  \rangle |_{z^2=0}
 } \label{eq:nfwddef}  \\ \hspace{-1cm} &&  
= \bar u(p-r)  \hat z 
 u(p) \, (z \cdot p) \int_0^1   \int_0^1  \, 
\frac1{2}  \left ( e^{-ix(pz)-iy(r z)} 
+e^{ix(pz)-i\bar y(r z)}\right ) 
\theta( x+y \leq 1) \,  F_g(x,y)  
 \,  dx \, dy .
\nonumber
\end{eqnarray} 
Here and in the following I adhere to  the convention
$\bar y = 1-y, \bar x = 1-x$, $etc.,$ 
 for  momentum fractions and use the notation
$\gamma_{\alpha} z^{\alpha} \equiv \hat z$.

Due to  the spectral properties $x \geq 0$,
$y \geq 0$, $x+y \leq 1$  (this can be proved for any Feynman 
diagram using the approach of ref.
\cite{spectral}), 
  both the initial active gluon 
and the spectators  carry  positive 
fractions of the initial hadron  momentum $p$:
$(x+\zeta y)$ for the gluon  and 
$(\bar x- \zeta y)  \geq y(1-\zeta) \geq 0$ 
for  the  spectators. 
On the other hand, the fraction of the $p$-momentum 
carried by another
gluon   is given by 
$(x - \bar y \zeta)$ and it may take  both 
positive and negative values.

The usual gluon  distribution function $xf_g(x)$
corresponds to the limit $r =0$.
Hence, the double  distribution $F_g(x,y)$ 
satisfies the  reduction formula:
\begin{equation}
\int_0^{1-x} \, F_g(x,y)\, dy= x f_g(x) \, .
\label{eq:redf}
\end{equation}

{\it 3. Asymmetric distribution  functions.}
Since $r = \zeta p$,  the variable $y$ appears 
in eq.(\ref{eq:nfwddef}) only in the combinations
$x+y\zeta \equiv X$ and $x- \bar y\zeta \equiv  X - \zeta$,
where $X$ and  $(X - \zeta)$ are   the total fractions 
of the initial hadron momentum $p$ carried by the  gluons
(cf. \cite{afs}).
Introducing $X$ as an independent variable, we  can 
integrate the  double   distribution $F(X-y \zeta,y)$ 
over $y$ to get
\begin{equation}
{\cal F}_{\zeta} (X) =  \int_0^{{\rm min} \{ X/\zeta, 
\bar X / \bar \zeta \}} F(X-y \zeta,y) \, dy, \label{eq:asdf}
\end{equation}
where $\bar \zeta \equiv 1- \zeta$.
Since $\zeta \leq 1$ and  $x+y \leq 1$, 
the variable $X$ satisfies a natural
constraint $0\leq X \leq 1$.

\begin{figure}[ht]
\mbox{
   \epsfxsize=12cm
 \epsfysize=5cm
 \hspace{2cm}  
 \epsffile{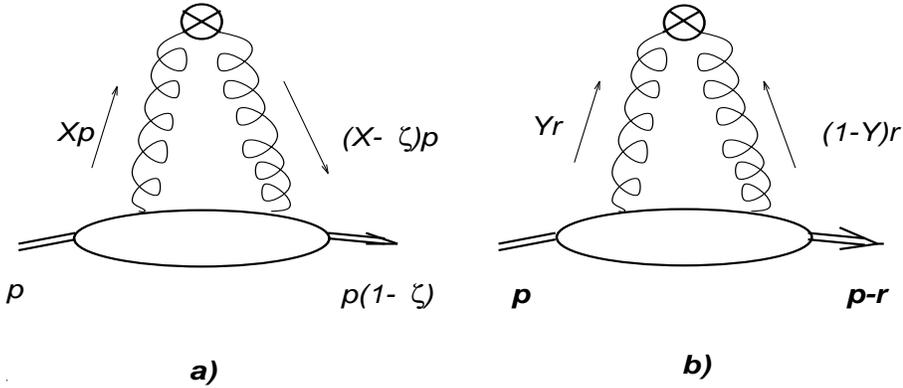}  }
{\caption{\label{fig:2}
Momentum flow corresponding to the  asymmetric gluon distribution 
function in two regimes:
{\it a) } $X \geq \zeta$ and {\it b)}
 $X \equiv Y \zeta \leq \zeta $.
   }
}
\end{figure}

In the region $X > \zeta$ (Fig.2$a$),
where the initial gluon momentum $Xp$ 
is larger than the momentum transfer $r = \zeta p$,
the function ${\cal F}_{\zeta}^g  (X)$ 
can be treated as  a generalization of the 
usual  distribution function $xf_g(x)$
for the asymmetric case when the final hadron momentum $p'$ 
differs by $\zeta p$ from the initial momentum $p$ (but remains
collinear to it).  
In this case,  ${\cal F}_{\zeta}^g  (X)$   describes
 a gluon   going out of 
the hadron with a positive fraction  $Xp$  
of the original hadron momentum
and then coming back into the hadron with a changed 
(but still positive) fraction  $(X - \zeta)p$.
The Bjorken ratio   $\zeta$ serves  here as an external
parameter specifying the momentum asymmetry  of 
the matrix element. Hence, one 
deals now  with a family of 
asymmetric distribution functions  ${\cal F}_{\zeta}^g (X)$ 
whose shape changes when $\zeta$ is changed.
The  basic distinction  between 
the  double  distributions $F(x,y)$ 
and the asymmetric  distribution functions 
${\cal F}_{\zeta} (X)$ is that   the former  
are universal functions which 
do not depend on the momentum asymmetry  parameter
$\zeta$, while the latter are explicitly labelled by it.
When $\zeta  \to 0$, the  limiting curve 
for ${\cal F}_{ \zeta}(X)$ reproduces the 
gluon  distribution function:
 \begin{equation}
 {\cal F}^g_{\zeta=0} \, (X) =  X f_g(X) \  . \label{eq:Fzeta0}
\end{equation}

Another region is $X < \zeta$ (Fig.2$b$), 
in which  the ``returning'' gluon has
a negative fraction $(X- \zeta)$ of the light-cone momentum $p$.
Hence, it is more appropriate to  treat it  as a gluon 
going out of the hadron and 
propagating  together with the original one.
Writing $X$ as $X = Y \zeta$, we immediately obtain that 
the gluons  carry now 
positive fractions $Y \zeta p \equiv Y r$ and,
respectively,  $(1-Y)r \equiv \bar Y r$ of 
the momentum transfer $r$. 
Hence, in the region $X= Y \zeta < \zeta$, the
asymmetric distribution function
looks like a distribution amplitude 
$\Psi_{\zeta}(Y)$ for a  two-gluon 
state with the  total momentum $r= \zeta p$: 
\begin{equation}
\Psi_{\zeta}(Y) =  \int_0^Y F((Y-y) \zeta , y ) \, dy . 
\label{eq:Psi}
\end{equation}

The   asymmetric  distribution function 
can be also defined directly 
through the matrix element
\begin{eqnarray} 
\hspace{-1cm} &&\lefteqn{\langle p'  \,  | \,   
z_{\mu}  z_{\nu} G_{\mu \alpha}^a (0) E_{ab}(0,z;A) 
G_{ \alpha \nu }^b (z)\, | \,p  \rangle |_{z^2=0}
 } \label{eq:asdef}  \\ \hspace{-1cm} &&  
= \bar u(p')  \hat z 
 u(p) \, (z \cdot p) \, \int_0^1   
\frac1{2}  \left ( e^{-iX(pz)}
 + e^{i(X-\zeta)(pz)} \right ) 
 {\cal F}^g_{\zeta}(X)  
 \,  dX .
\nonumber
\end{eqnarray} 
To re-obtain the  relation between ${\cal F}_{\zeta}(X)$
and  the double distribution 
function $F_g(x,y)$, one should combine this definition
with eq.(\ref{eq:nfwddef}).

{\it 4. Leading contribution.} 
The parameterization 
(\ref{eq:asdef})  can be used as  a starting point
for constructing   a QCD parton-type formalism.
The only problem is that our gauge-invariant
definition of the gluon distribution
is in terms of  the field strength tensor
$G_{\mu \nu}$ while the usual Feynman rules
involve  the vector potential $A_{\mu}$. 
A possible way out is to utilize  the light-cone gauge
$q^{\mu} A_{\mu} (z;q) = 0$ in which    $A_{\mu}$
can be expressed in terms of  $G_{\mu \nu}$
\begin{equation} 
A_{\mu}(z;q) = q^{\nu} \int_0^{\infty} 
G_{\mu \nu}(z+\sigma q) \, d \sigma , \label{eq:lcgauge}  
\end{equation}
so that the definition (\ref{eq:asdef}) 
can be applied directly. 
This gives
\begin{eqnarray} 
\hspace{-1cm} &&\lefteqn{\langle p' \,  | \,   
A_{\mu}^a (z_1;q)
A_{\nu}^a (z_2;q)\, | \,p  \rangle |_{z_1^2=0,z_2^2=0}
=  \frac {\bar u(p')  \hat q 
 u(p)}{2 (q \cdot p)}
 \, \biggl (- g_{\mu \nu} + 
\frac{p_{\mu}q_{\nu}+p_{\nu} q_{\mu}}{(p \cdot q)} \biggr ) 
} \label{eq:AAdef}  \\ &&  
\times \int_0^1   \left ( e^{-iX(pz_1)+i(X-\zeta)(pz_2)}
 + e^{i(X-\zeta)(pz_1)-iX(pz_2)} \right ) 
 \frac{{\cal F}^g_{\zeta}(X)}{X(X- \zeta +i \epsilon)}  
 \,  dX .
\nonumber 
\end{eqnarray} 

In ref.\cite{bfgms},   the amplitude 
of hard diffractive electroproduction was calculated 
for the longitudinal polarization of both the 
virtual photon ($\epsilon_{\gamma^*}^{\mu} = 
(q^{\mu}+ \zeta p^{\mu})/Q$)
and produced meson ($\epsilon_{V}^{\mu} =q^{\mu}/m_V$).
In this case,  we obtain 
\begin{equation} 
T_{LL} (p,q,r)  \sim   \frac{\sqrt{1-\zeta}}{Q m_V} \int_0^1 
\frac{\varphi_{V}(\tau)}
{\tau \bar \tau} \, d \tau
\int_0^{1} 
  \frac{{\cal F}^g_{\zeta}(X) dX}{X(X-\zeta+i\epsilon)} \, , 
\label{eq:tLL}
\end{equation}
where $\sqrt{1-\zeta}$  comes from 
$\bar u(p')  = \sqrt{1-\zeta} \, \bar u(p)$
and $\varphi_V(\tau)$ 
is the distribution amplitude of the longitudinal 
 vector meson. 
The  ampitude 
has the imaginary part due to the factor  $1/(X-\zeta +i\epsilon)$:
\begin{eqnarray} 
 \frac1{\pi}\, {\rm Im} \, T_{LL}(\zeta ) 
\sim    \frac{\sqrt{1-\zeta} }{\zeta Q m_V}
\, {\cal F}^g_{\zeta} \, (\zeta)\, 
\int_0^1 \frac{\varphi_V(\tau)}
{\tau \bar \tau} \, d \tau  \, .
\label{eq:imtLL}
\end{eqnarray} 
In ref.\cite{bfgms},
the gluonic matrix element was approximated by the 
gluon distribution function $f_g(\zeta)$. 
To get our result from that of ref.\cite{bfgms},
one should substitute there $f_g(\zeta)$ by  $\sqrt{1-\zeta} \,
{\cal F}^g_{\zeta}(\zeta) /\zeta$.

Though 
the asymmetric  distribution function 
${\cal F}^g_{\zeta}(X)$ coincides with 
$X f_g(X)$ in the limit $\zeta =0 $,  
these two functions differ in the general
case when $\zeta \neq 0$.
Furthermore,  the imaginary part 
appears for $X= \zeta$, $i.e.,$ in a
highly asymmetric  configuration in which the second gluon 
carries a vanishing fraction
of the original hadron momentum.
Hence,  one cannot exclude  the possibility  that 
${\cal F}^g_{\zeta}(\zeta)$
visibly   differs from the function 
$\zeta f_g(\zeta)$ which corresponds to a symmetric 
 configuration  in which the final gluon
has  the momentum equal to that of  the initial one.

To  get a feeling about the interrelationship 
between ${\cal F}^g_{\zeta}(X)$ 
and $X f_g(X)$,  let us consider a  toy
model 
$
F^{mod}(x,y) = A (n+1) (1-x-y)^n
$
for the gluon double  distribution.
Then  $x f^{mod}(x) =  A (1-x)^{n+1} $ while 
\begin{equation}
{\cal F}^{mod}_{\zeta}(X) =  \frac{A}{1-\zeta}
\biggl [ \left ( (1-X)^{n+1} - (1-X / \zeta)^{n+1} \right ) 
\theta (X < \zeta) + (1-X)^{n+1}\theta (X > \zeta) \biggr ] \ . 
\label{eq:Fmod}
\end{equation}
Hence, $\zeta f^{mod}(\zeta) = A (1-\zeta)^{n+1}$
while ${\cal F}^{mod}_{\zeta}(\zeta) = A (1-\zeta)^{n}$,  $i.e.,$
the two functions are rather close to each other for
small $\zeta$. This model also reveals 
a characteristic feature of the asymmetric distribution
function ${\cal F}^{mod}_{\zeta}(X)$:
the  parameter  $\zeta$ specifying the 
momentum asymmetry of the gluonic matrix element
serves as a boundary between the two regions
$X < \zeta$ and $X > \zeta$
in which  ${\cal F}^{g}_{\zeta}(X)$
is given by different analytic expressions.
An important property of ${\cal F}^{mod}_{\zeta}(X)$
is that it rapidly  varies  in the region $X \lapprox \, \zeta$ 
and  vanishes for $X=0$.
Since $X=0$ can be arranged only when  both $x$- and $y$-parameters 
of the double  distribution $F(x,y)$
are set to zero, the  fact that  ${\cal F}^{mod}_{\zeta}(0) =0$
is  quite general.
Note, however, that the limiting curve ${\cal F}^{mod}_{\zeta=0}(X)=
(1-X)^{n+1}$ does not vanish for $X=0$,
$i.e.,$ the limits $\zeta \to 0$ and $X \to 0$ do not commute.

For this reason,  if $\zeta$ is small, the substitution  of 
${\cal F}_{\zeta}(X)$ by $X f_g(X)$ may be a good approximation
for all $X$-values except for the region $X \lapprox \, \zeta$.
However, the imaginary part is given  by the value of 
 ${\cal F}_{\zeta}(X)$ at the point  $X=\zeta$
belonging to this region and it is not 
clear {\it a priori}  how close are the functions
 ${\cal F}_{\zeta}(\zeta)$ and $\zeta f_g(\zeta)$.

{\it 5. Evolution of the double distribution.} 
On the light-cone,  the matrix
elements  have ultraviolet divergences,
which are removed by 
subtraction  prescription characterized 
by  a scale $\mu$: 
$F_g(x,y) \to F_g(x,y;\mu)$.
Under renormalization, the gluonic 
operator 
\begin{equation}
{\cal O}_g(uz,vz) =
z_{\mu}  z_{\nu} 
G_{\mu \alpha}^a (uz) E_{ab}(uz,vz;A) 
G_{\nu \alpha}^b (vz) \label{eq:Og}
 \end{equation}
mixes with the flavor-singlet 
quark operator
\begin{equation}
{\cal O}_Q(uz,vz) = \frac{i}{2} \sum_q (\bar \psi_q(uz) 
\hat z E(uz,vz;A)  \psi_q(vz)
- \bar \psi_q(vz) \hat z  E(vz,uz;A) \psi_q(uz))\, . \label{eq:OQ}
 \end{equation}

For simplicity, we will 
ignore here the quark-gluon mixing and 
analyze  below 
the evolution of the double gluon 
distribution $\tilde F_g(x,y;\mu)$
corresponding to the ``quenched approximation''. 
In this case 
\begin{equation}
 \mu \frac{d}{d \mu}  \tilde F_g(x,y;\mu) =
\int_0^1 d \xi \int_0^1  R_{gg}(x,y; \xi, \eta;g(\mu)) 
\tilde F_g( \xi, \eta;\mu) d \eta.
\label{eq:nfwdev} 
\end{equation}

The easiest way to get  explicit expressions 
for $R_{ab}(x,y; \xi, \eta;g)$  is 
to use the Balitsky-Braun 
evolution equation \cite{bb}
for the light-cone 
operators\footnote{Instead of the  original kernels
 $K_{ab}( u,v )$ from ref.\cite{bb} 
we prefer to use  the kernels
$B_{ab}( u,v ) = -K_{ab}(\bar u,v ) $
which have the symmetry property 
$B_{ab}( u,v ) = B_{ab}( v,u )$  .}
\begin{equation}
 \mu \frac{d}{d \mu} 
{\cal O}_a(0,z)    =
\int_0^1  \int_0^{1}  
\sum_{b} B_{ab}(u,v ) {\cal O}_b( uz, \bar vz) \,
\theta (u+v \leq 1) \, du \, d v  \,  , 
\label{eq:balbr} 
\end{equation}
where $a,b = g,Q$ and  \cite{bb}  
\begin{eqnarray}
B_{gg}(u,v ) = \frac{\alpha_s}{\pi} N_c \biggl (
4(1 + 3uv -u -v) + \frac{\beta_0}{2 N_c} \,
\delta(u)\delta(v) 
\nonumber \\ 
+ \delta(u) \biggl[ \frac{\bar v^2}{v} - \delta (v) \int_0^1
\frac{d z}{z} \biggr ] 
+ \delta(v) \biggl[ \frac{\bar u^2}{u} - \delta (u) \int_0^1
\frac{d z}{z} \biggr ] 
 \biggr  )  \, .
\label{eq:Bgg}
\end{eqnarray}
Here,  $\beta_0 = 11- \frac2{3} N_f$ 
is the lowest coefficient of the 
QCD $\beta$-function.

Our  kernel $R_{gg}(x,y; \xi, \eta;g)$ is  related to 
the $B_{gg}(u,v)$-kernel by
\begin{equation}
 R_{gg}(x,y; \xi, \eta;g) = \frac1{\xi} 
B_{gg} ( y - \eta x/\xi, \bar y - \bar \eta x/\xi).
\label{eq:RtoB}
\end{equation}
This gives 
\begin{eqnarray}
&& \hspace{-1cm} R_{gg}(x,y;\xi, \eta;g) = 
\frac{\alpha_s}{\pi} N_c \frac1{\xi} 
\Biggl  \{
4 [x/\xi + 3 (y-\eta x/\xi)(\bar y - 
\bar \eta x/\xi) ] \, 
 \theta  (0 \leq x/\xi \leq 
{\rm min} \{ y/\eta, \bar y / \bar \eta \} )
  \label{eq:rkernel}   \\
 && \hspace{-1cm}  +  
\frac{\theta (0 \leq x/\xi \leq 1) (x/\xi)^2}{ (1-x/\xi)} 
\left [ \frac1{\eta} \, \delta(x/\xi - y/\eta) + 
\frac1{\bar \eta} \, \delta(x/\xi - \bar y/ \bar \eta) \right]
+ \delta(1-x/\xi) \delta(y-\eta)
\left [ \frac{\beta_0}{2 N_c} -  \int_0^1 
\frac{dz}{z} \right ]\, \Biggr \}.
\nonumber
\end{eqnarray}
As usual,  the  divergent integral  
provides the regularization for the singularities 
of the kernel for  $x=\xi$ (or $y= \eta$).
It is easy to verify that the kernel
$R_{gg}(x,y;\xi, \eta;g)$ has the property that 
$x+y \leq 1$ if $\xi+ \eta \leq 1$. 
Using the   expression for $R_{gg}(x,y;\xi, \eta;g)$ and the explicit
form of the GLAPD kernel $P_{gg}(x/\xi)$  \cite{ap,d}, 
 one can  check the reduction formula
 \begin{equation}
\int_0^ {1-x}  R_{gg}(x,y; \xi, \eta;g) d y =
\frac1{\xi} P_{gg}(x/\xi).
\label{eq:rtop} 
\end{equation}
Integrating $R_{gg}(x,y; \xi, \eta;g)$
over $x$ one should get  the  evolution kernel
for the gluon distribution amplitude 
\begin{equation}
\int_0^{1-y}   R_{gg}(x,y; \xi, \eta;g) d x = V_{gg}(y,\eta;g).
\label{eq:rtov}
\end{equation}

To solve the evolution equation, we 
apply first the  standard trick used to solve the GLAP equation:
integrate  $x^n R_{gg}(x,y; \xi, \eta;g)$ over $x$. 
Using the property 
 $R_{gg}(x,y; \xi, \eta;g) = R_{gg}(x/\xi,y; 1, \eta;g)/\xi$,
we get  
\begin{equation}
 \mu \frac{d}{d \mu} \tilde F^{(n)}_{g} (y;\mu) =
\int_0^1  R^{(n)}_{gg} (y,\eta;g) 
\tilde F^{(n)}_{g}(\eta;\mu) d \eta \, ,
\label{eq:fnev}
\end{equation}
where  $\tilde F^{(n)}_{g}(y;\mu)$ is the $n$th $x$-moment of $\tilde F_g(x,y;\mu)$
\begin{equation}
\tilde F^{(n)}_{g}(y;\mu) = \int_0^{1} x^n  \tilde F_{g}(x,y;\mu) dx
\label{eq:fnmom}
\end{equation}
and the kernel $R^{(n)}_{gg} (y,\eta;g)$  is given by
\begin{eqnarray}
&& \lefteqn{R^{(n)}_{gg} (y,\eta;g) = 
\frac{\alpha_s}{\pi} N_c 
\left \{   \left (\frac{y}{\eta} \right )^{n+2} 
\left ( \frac4{n+2} +\frac{12}{n+1}\bar y \eta - \frac{12}{n+2}
(y \bar \eta + \eta \bar y) + \frac{12}{n+3} y \bar \eta 
\right. \right. } \nonumber \\
&& +  \left.  \left.
 \frac{1}{\eta -y} \right ) \theta(y \leq \eta)
 + \delta(y-\eta) 
\left [ \frac{\beta_0}{2 N_c} - \int_0^1 \frac{dz}{z} \,  \right ]
+ \{y \to \bar y, \eta \to \bar \eta \} \right \}.
\label{eq:rnkernel}
\end{eqnarray}
It is straightforward to establish 
that  $R^{(n)}_{gg}  (y,\eta;g)$ has the property
$$R^{(n)}_{gg}  (y,\eta;g) w_n(\eta) =
R^{(n)}_{gg}  (\eta,y;g) w_n(y) ,$$ where $w_n(y)= (y \bar y)^{n+2}$. 
Hence, the  eigenfunctions of $R^{(n)}_{gg}  (y,\eta;g)$ 
are orthogonal with
the weight $w_n(y)= (y \bar y)^{n+2}$, $i.e.,$ 
they are proportional to the Gegenbauer polynomials
$C^{n+5/2}_k(y-\bar y)$ (cf.\cite{bl,mikhrad} and 
refs.\cite{shvys,ohrn} where
the general algorithm was originally applied to the 
evolution of gluonic distribution amplitudes).
Now, we can write  the general solution of the evolution
equation
\begin{equation} 
\tilde F^{(n)}_{g}(y;\mu) = (y \bar y)^{n+2} \sum_{k=0}^{\infty} A_{nk}
C^{n+5/2}_k(y-\bar y) \left [\log (\mu /\Lambda) \right]^
{-\gamma^{(n)}_k/\beta_0},
\label{eq:fnsol}
\end{equation}
where  the anomalous dimensions $\gamma^{(n)}_k$ 
 are given by 
the eigenvalues of the kernel
$ R^{(n)}_{gg}  (y,\eta;g)$:
\begin{equation}
\gamma^{(n)}_k= 2 N_c \left [- \frac1{(k+n)(k+n+1)}  
- \frac1{(k+n+2)(k+n+3)}
 + \sum_{j=1}^{k+n+1} \frac1{j}\right ] 
 -\frac1{2} \beta_0.
\label{eq:ads}
\end{equation}
They coincide with the standard 
anomalous dimensions $\gamma_{gg}(N)$ \cite{gw,gp}:  
$\gamma^{(n)}_k = \gamma_{gg}(n+k+1)$. 
Note, that $\gamma_0^{0}$  is formally given by 
the negative infinity, while  all other anomalous  dimensions 
are finite and  non-negative. Hence,  in the formal $\mu \to \infty$
limit we have  
$\tilde F_n(y, \mu \to \infty) =0$  for all $n \geq 1$.
This means that 
$$\tilde F_g(x,y; \mu \to \infty)  \sim \delta(x) (y \bar y)^2,$$
$i.e.,$ in each of its variables the limiting function 
 $\tilde F_g(x,y; \mu \to \infty)$  
acquires the characteristic asymptotic form dictated by
the nature of the variable:
$\delta(x)$ is specific for the distribution functions \cite{gw,gp},
while  the $(y \bar y)^2$-form  is  
the asymptotic shape   for the lowest-twist gluonic 
distribution amplitudes \cite{shvys,ohrn}.
It is easy to see that if  the double distribution 
has the asymptotic form  
$ F^{as}(x,y) =  120 \, C \, \delta(x) \, y^2 (1-y)^2$,
then   $x f^{as}(x) = C \, \delta(x)$
while ${\cal F}^{as}_{\zeta}(X) = 
120 \, C \, X^2 (1-X/\zeta)^2 /\zeta $.
Note that in this case ${\cal F}^{as}_{\zeta}(\zeta) =0$,
$i.e.,$ the function which determines the magnitude of the imaginary
part of $T_{LL}$ vanishes.

{\it 6. Evolution equations for asymmetric distribution functions.}
Introducing the asymmetric distribution function
${\cal F}_{\zeta}^Q (X)$ for the 
flavor-singlet quark combination (\ref{eq:OQ})
\begin{equation} 
\langle p-r\, | \, {\cal O}_Q(0,z)\, | \, p \rangle |_{z^2=0}  = 
i \bar u(p-r)  \hat z 
 u(p)   \int_0^1    \, 
  \left ( e^{-iX(pz)} - e^{i(X-\zeta)(pz)}\right )
{\cal F}_{\zeta}^Q (X) 
 \,  dX
\label{eq:Q}
\end{equation}
and using eq.(\ref{eq:balbr}), we obtain 
a set of coupled evolution equations 
for ${\cal F}^g_{\zeta}(X)$ and ${\cal F}^Q_{\zeta}(X)$:
\begin{equation}
 \mu \frac{d}{d\mu}  {\cal F}_{\zeta}^a(X;\mu) =
\int_0^1  \, \sum_b \,  W_{\zeta}^{ab}(X,Z;g) \, 
{\cal F}_{\zeta}^b( Z;\mu) \, d Z \,  ,
\label{eq:asev} 
\end{equation}
where $a$ and $b$ denote $g$ or $Q$.
To relate the functions  $ W_{\zeta}^{ab}(X,Z;g)$ 
to the Balitsky-Braun  evolution kernels \cite{bb}
\begin{eqnarray}
B_{QQ}(u,v ) = \frac{\alpha_s}{\pi} C_F \left 
(1 + \delta( u) [\bar v/v]_+  + 
\delta(v) [\bar u/ u]_+ - \frac1{2} \delta( u)\delta(v) \right ) \, ,
\label{eq:BQQ} 
\\
B_{gQ}(u,v ) = \frac{\alpha_s}{\pi} C_F \biggl 
(2 + \delta( u)\delta(v) \biggr  )
 \ \ ,  \  \
B_{Qg}(u,v ) = \frac{\alpha_s}{\pi} N_f \left 
(1 + 4uv -u -v \right ) \label{eq:BQG} 
\end{eqnarray}
($B_{gg}(u,v)$ was displayed  earlier by  eq.(\ref{eq:Bgg})),
it is convenient to  introduce first the auxiliary 
kernels $ M^{ab}_{\zeta}(X,Z;g)$:
\begin{equation}
M^{ab}_{\zeta}(X,Z) = \int_0^1  \int_0^1 B_{ab}(u,v) \, 
\delta(X- \bar u Z + v (Z- \zeta)) \,
\theta(u+v \leq 1)
\, du \, dv  \, . 
\label{eq:M}
\end{equation}
In terms of these kernels, we have 
\begin{eqnarray}
W^{gg}_{\zeta}(X,Z)=M^{gg}_{\zeta}(X,Z) \  \    ,   \  \ 
W^{QQ}_{\zeta}(X,Z)=M^{QQ}_{\zeta}(X,Z), \\ 
W^{gQ}_{\zeta}(X,Z)= \int_X^1 M^{gQ}_{\zeta}(\widetilde X,Z)
\, d \widetilde X \ \  , \  \
W^{Qg}_{\zeta}(X,Z)=\frac{d}{dX} \, M^{Qg}_{\zeta}(X,Z) \, .
\end{eqnarray}
Note that the expressions
for the  kernels describing the quark-gluon  mixing 
are slightly more involved than those for the diagonal ones.
The reason is that the definition (\ref{eq:asdef}) of 
the gluon distribution has an extra $(z \cdot p)$ factor which 
generates  a derivative acting on ${\cal F}_{\zeta}^g(X)$.

Integrating  the 
 delta-function in eq.(\ref{eq:M}), 
one obtains  four different types of the $\theta$-functions,
each of which 
corresponds to a specific
evolution regime for  the asymmetric 
distribution functions. 
In particular, for $W_{\zeta}^{gg}(X,Z;g)$ we obtain 
 \begin{eqnarray}
&& \lefteqn{ W_{\zeta}^{gg}(X,Z;g) = \frac1{Z} \int_0^1  dv \, 
B_{gg} \biggl  ( [1- X/Z -v(1-\zeta/Z)] \, ,v \biggr ) }
 \label{eq:Wgg} \\
&& \times \biggl \{ 
\, \theta(Z\geq X \geq \zeta ) 
\, \theta \left (0 \leq v \leq \frac{1-X/Z}{1- \zeta/Z} \right )
+\, \theta(Z \geq \zeta \geq X ) \, \theta(0 \leq v \leq X/ \zeta ) 
\nonumber \\ && +  \theta (X \leq \zeta ) \, \theta(Z \leq \zeta )  
 \left [ \, 
\, \theta (X \leq Z) \, \theta(0 \leq v \leq X/ \zeta ) 
+
\theta(X \geq Z) \, \theta \left ( \frac{X/Z-1}{\zeta/Z -1 } 
\leq v \leq X/\zeta \right )
\right  ]
\biggr  \} .
\nonumber
 \end{eqnarray}

For the first two terms in this sum,  the original fraction 
$Z$ is bigger than the momentum asymmetry parameter $\zeta$. 
In this region of $Z$'s, the  resulting fraction $X$ 
cannot  be increased by the evolution:
in both terms we have $X \leq Z$. Such a situation is typical
for the evolution of distribution functions.
 
The last two terms in eq.(\ref{eq:Wgg}) correspond to the region
where the original fraction 
$Z$ is smaller  than  $\zeta$. 
In this case, the evolution can either decrease the fraction
($X \leq Z$ for the first term in the square brackets) or increase it 
( $X \geq Z$ in the second one). 
However, if the initial fraction 
$Z$  is less than $\zeta$, the evolved fraction 
$X$ cannot be larger than $\zeta$ or, what
is the same, the parameter $Y \equiv X/\zeta$
specifying the fraction $Yr$ of the momentum transfer $r$
carried by this  gluon  cannot exceed $1$. 
This property is a characteristic  feature of the evolution
of distribution amplitudes.

Qualitatively,  the evolution of the asymmetric distribution functions
proceeds in the following way.
Due to the GLAP-type evolution, 
the momenta of the partons decrease, and distributions
become peaked in the regions of smaller
and smaller $X$. However, when the parton momentum
degrades  to values smaller than the momentum transfer
$r = \zeta p$, the further evolution is like 
that for a distribution amplitude:
it tends to make the distribution symmetric with respect to
the central point $X= \zeta/2$ of the $(0, \zeta)$ segment. 
In two extreme cases,  when
 $\zeta =0$ or $\zeta =1$, the evolution is more 
trivial. For $\zeta =0$, 
 ${\cal F}_{\zeta}(X)$ reduces to a usual
distribution function governed by the GLAP
evolution equation, while for 
$\zeta =1$ we always have $Z \leq \zeta$ 
for the initial fraction
and the function ${\cal F}_{\zeta}(X)$ 
experiences  a purely Brodsky-Lepage evolution. 
In other words, 
$W_{\zeta =0}^{ab}(X,Z;g) = P_{ab}(X/Z,g)/Z$ and 
$W_{\zeta =1}^{ab}(X,Z;g) =V_{ab}(X,Z;g)$
\footnote{
Originally, this  observation was  made by
X.Ji and P.Hoodbhoy \cite{jil} in application 
to  evolution  of the 
$t=0$ limit of the off-forward parton distributions  
introduced in ref.\cite{ji}. }.

{\it 7. Conclusions.} In this letter, I demonstrated that
one can describe the asymmetric matrix
element $\langle  \bar \zeta p | G \ldots G |p \rangle$
either by the  universal double distribution
$F_g(x,y)$ or by the asymmetric distribution
function ${\cal F}_{\zeta}(X)$  which explicitly 
depends on the momentum asymmetry parameter $\zeta$ 
and specifies the total fractions $X$ and $X - \zeta$ 
of the original hadron momentum $p$ carried  
by the gluons. Using ${\cal F}_{\zeta}(X)$ 
gives a formalism that looks very similar to the standard 
QCD parton approach, in which  the gluon 
content of the hadron is described by the gluon
distribution function $f_g(x)$. 
Moreover, ${\cal F}_{\zeta}(X)$  coincides with 
$X f_g(X)$ in the $\zeta \to 0$ limit,
and this fact suggests the approximation
${\cal F}_{\zeta}(X) \approx X f_g(X)$ for small $\zeta$.
One should realize, however, that 
the electroproduction amplitude is dominated 
by the imaginary part whose magnitude is determined by
${\cal F}_{\zeta}(\zeta)$. 
Since the function ${\cal F}_{\zeta}(X)$ rapidly varies 
in the region $X \lapprox \, \zeta$ and vanishes
for $X=0$ (which is not the case with
$X f_g(X)$), the relation ${\cal F}_{\zeta}(\zeta) 
\approx \zeta f_g (\zeta)$
may be strongly violated. This 
pessimistic expectation  is not supported 
by a toy model for the double distribution
$F^{mod}(x,y) = A (1-x-y)^n$,
in which  $\zeta f_g^{mod} (\zeta)$ differs from 
 ${\cal F}_{\zeta}^{mod}(\zeta)$
by an extra factor $(1- \zeta)$ only, the latter
being close to 1 for small $\zeta$.
However,  the structure of  
the double distribution  $F_g(x,y)$ 
in general case may be more involved,
and a detailed analysis  of the interrelationship between
${\cal F}_{\zeta}(\zeta)$ and $\zeta f_g (\zeta)$ 
is an interesting problem
for  future studies.

{\it Acknowledgements.}
I am grateful to  Mark Strikman 
for  discussions which stimulated 
this investigation and  to Xiangdong Ji 
for correspondence and criticism concerning ref.\cite{compton}. 
I thank Ian Balitsky and Igor Musatov for useful discussions
and  Leonid Frankfurt for comments.
My special gratitude is to Nathan  Isgur for continued 
encouragement and advice.
This work is supported
by the US Department of Energy 
under contract DE-AC05-84ER40150.
I thank the DOE's Institute for Nuclear Theory
at the University of Washington for its hospitality
and support  during the workshop
``Quark and Gluon  Structure of Nucleons and Nuclei'' 
where this work has been started.

\end{document}